# Sub-kHz single-frequency pulsed semiconductor laser based on NPRO injection locking

Chunzhao Ma,[a] Wenxun Li,[a] Weitong Fan,[a] Danqing Liu,[a] Xuezhen Gong,[a] Zelong Huang,[a] Jie Xu,[a] Hsien-Chi Yeh[a], And Changlei Guo[a,*]

[a] MOE Key Laboratory of TianQin Mission, TianQin Research Center for Gravitational Physics & School of Physics and Astronomy, Frontiers Science Center for TianQin, CNSA Research Center for Gravitational Waves, Sun Yat-sen University (Zhuhai Campus), Zhuhai 519082, China

A B S T R A C T

We report a single-frequency, narrow-linewidth semiconductor pulsed laser based on pump current modulation and optical injection locking technique. A monolithic non-planar ring oscillator laser is employed as the seed source to guarantee the single-frequency narrow-linewidth performance. Simultaneously, pulse operation is achieved by directly modulating the pump current of the semiconductor laser. The single-frequency pulsed laser (SFPL) has achieved a pulse repetition rate of 50 kHz-1 MHz, a pulse duration ranging from 120 ns to a quasi-continuous state, and a peak power of 160 mW. Moreover, the SFPL has reached a pulsed laser linewidth as narrow as 905 Hz, optical spectrum signal-to-noise ratio of better than 65 dB at a center wavelength of 1064.45 nm. Such extremely narrow-linewidth, repetition-rate and pulse-width tunable SFPL has great potential for applications in coherent LIDAR, metrology, remote sensing, and nonlinear frequency conversion.

## 1. Introduction

In recent years, single-frequency pulsed lasers (SFPL) have found widespread applications, such as coherent Doppler LIDAR [1–5], remote sensing [6,7], metrology [8,9], and nonlinear frequency conversion [10]. To meet the requirements of these applications, a great deal of pioneering and innovative work has been carried out to improve the operation performance of SFPLs, especially in laser linewidth compression. A common approach to generate SFPL is to externally modulating a continuous-wave (CW) laser operating at a single frequency [11–14]. While this method is relatively straightforward to implement, it is associated with inherent power losses. To overcome the shortcoming of external modulation, one typical approach is to modulate the intra-cavity losses of lasers, also known as *Q*-switching. By inserting an electro-optic or acousto-optic modulator into a ring cavity and injecting a narrow-linewidth seed laser, researchers have successfully produced a single-frequency narrow-linewidth *Q*-switched pulsed laser output [15–17]. In addition, an all-fiber single-frequency *Q*-switched laser was achieved by polarization modulation through stress-induced birefringence using a piezoelectric transducer (PZT) as a *Q*-switcher [18,19]. Besides the methods of actively adjusting intra-cavity losses, Li et al. passively *Q*-switch a fiber laser using a saturation absorber and used an unpumped EDF fiber together with a fiber Bragg grating to select the longitudinal mode, thus realizing a single-frequency, narrow-linewidth pulsed laser output [20].

*Q*-switching, though widely used in pulsed laser generation, introduces technical complexities and causes extra power losses due to added cavity components. Gain-switching offers another solution. By modulating the pump source, pulsed laser output with various parameters can be achieved without extra components. Geng et al. achieved SFPL at 2.05 μm by gain-switching a Ho-doped laser with a 1.95 μm Tm-doped pulsed laser [21]. Hou et al. obtained pulsed output at 1.063 μm by pumping an Yb-doped fiber using a modulated laser diode (LD) [22]. Poozesh et al. employed the dynamic self-induced grating effect in Yb-doped fiber to generate a SFPL using a gain-switched pump source [23]. Fang et al. generated SFPL at 2.0 μm through the process of co-band pumping a DBR ultrashort cavity with a pulsed laser [24]. Zhang et al. employed pulsed pumping on a DFB ultrashort cavity and used pressure-induced birefringent phase shift to achieve SFPL [25]. Although gain-switching is relatively straightforward to implement, the pulsed laser linewidth typically falls within the megahertz range, which cannot meet stringent requirements of kilohertz linewidth for long-range detection [26]. Thus, it is essential to further compress the linewidth.

In the previously discussed implementation scheme for SFPLs, monolithic non-planar ring oscillator (NPRO) laser is an ideal choice as a seed source for a *Q*-switched laser, thanks to its single-frequency and narrow-linewidth characteristics. Nevertheless, the installation, debugging, and control of *Q*-switching components for monolithic NPRO laser raises a number of challenges [27–30]. On the other hand, semiconductor lasers exhibit a greater pulsed modulation bandwidth (up to gigahertz) compared to NPRO lasers. Nevertheless, such lasers typically require additional components for mode filtering and linewidth compression to achieve SFPL output [31–33]. The combination of the single-frequency, narrow-linewidth characteristics of NPRO lasers with the high-speed modulation capabilities of semiconductor lasers may facilitate a more straightforward implementation scheme for SFPLs with extremely narrow-linewidth.

In this paper, we experimentally demonstrate a sub-kHz linewidth SFPL at 1064 nm based on pump current modulation of a semiconductor laser that is injection locked to an NPRO seed. We first introduce the experimental setups for injection locking a distributed feedback laser diode (DFB-LD) to an NPRO seed, pulse generation with current modulation of the DFB-LD, and laser linewidth and pulse characterization of the SFPL. Experimental results show that, in the CW mode, the DFB-LD can output 100 mW 1064 nm laser with laser linewidth compressed from megahertz to sub-kHz after injection locking. In the pulsed mode, the DFB-LD has maintained the sub-kHz laser linewidth, and in the meantime, has achieved a pulse repetition rate of 50 kHz-1 MHz, a pulse duration ranging from 120 ns to quasi-continuous state and a peak power of 160 mW. We anticipate that such sub-kHz linewidth SFPL may find important applications in coherent LIDAR, metrology, remote sensing and nonlinear frequency conversion in the future.

## 2. Experimental setups

The schematic diagram of the single-frequency narrow-linewidth semiconductor pulsed laser is depicted in Fig. 1. The seed source is a homemade NPRO laser (NPRO1), which has a central wavelength of approximately 1064.45 nm. This single-frequency seed source is injected into a commercially available DFB-LD (Coherent, CMDFB1064B) through a fiber circulator (CIR). An optical isolator (ISO) is utilized to prevent the seed source from being influenced by the output of DFB-LD, thereby ensuring the integrity of the seed source is preserved. The pump current of the DFB-LD is pulse-modulated by a function generator (Rigol, DG4062) for the purpose of outputting a pulsed laser. Another homemade NPRO laser (NPRO2) serves as a reference laser and is combined with the output of the DFB-LD to form a heterodyne interferometer, which is used for subsequent frequency spectrum and frequency noise measurements. All three lasers are equipped with thermal electronic coolers (TEC) for temperature control and temperature-frequency tuning.

The optical signals from the DFB-LD and NPRO2 are divided into three paths through two optical couplers, as illustrated in Fig. 1. The first path is connected to an optical spectrum analyzer (Yokogawa, AQ6373B) to measure the optical spectrum. The second path is coupled to a scanning Fabry-Pérot interferometer (Thorlabs, SA210-8B) to measure the longitudinal modes or connected to a DC-coupled low-speed photodetector (PD2, Keyang Photonics, KY-PRM-200 M) to measure the pulse waveform and pulse repetition rates. The third path is connected to an AC-coupled high-speed photodiode (PD1, Keyang Photonics, KY-PRM-40 G) for characterizing the beat signal between the DFB-LD and NPRO2. The longitudinal modes or the pulse waveform are tested by an oscilloscope (Tektronix, MSO44). The frequency spectrum and frequency noise of the beat signal are measured by an electrical spectrum analyzer (R&S, FSW26) and a phase noise analyzer (R&S, FSWP26), respectively.

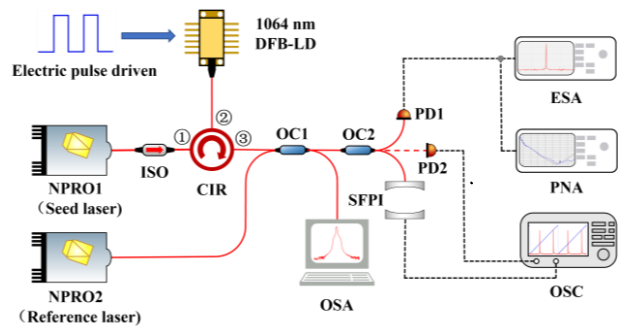

**Fig. 1** Schematic of the experimental setup. The DFB-LD is injection-locked to the seed laser NPRO1. The reference laser NPRO2 is used for beat signal generation. DFB-LD, distributed feedback laser diode; NPRO, non-planar ring oscillator; ISO, isolator; CIR, circulator; OC, optical coupler; OSA, optical spectrum analyzer; PD, photodetector; SFPI, scanning Fabry-Pérot interferometer; ESA, electrical spectrum analyzer; PNA, phase noise analyzer; OSC, oscilloscope.

## 3. Results and discussion

To investigate the influence of injection locking on the linewidth compression of the DFB-LD, we first assessed the linewidth of the DFB-LD under continuous operation. Before initiating the experiment, the frequency detuning between NPRO1 and NPRO2 was tuned to about 4.6 GHz to ensure that the electrical spectrum analyzer and the phase noise analyzer could detect the beat signal. In order to achieve injection locking of the DFB-LD with NPRO1, it is essential to adjust the frequency of either NPRO1 or the DFB-LD to ensure that the frequency detuning between the two lasers falls within the locking bandwidth [34]. However, the temperature-frequency tuning range of the NPRO laser (20 GHz) is relatively low in comparison with that of the DFB-LD (100s GHz). Furthermore, when the temperature of the NPRO is adjusted, phenomena like mode-hopping or dual-frequency lasing may occur, which may have a negative impact on injection locking [35]. To avoid this issue, we will adjust the temperature set point of the DFB-LD in subsequent experiments, thereby modifying the frequency detuning between the two lasers.

Fig. 2 shows the beat signal (between the DFB-LD and NPRO2) characterizations before and after the DFB-LD is injection locked. In Fig. 2(a), one can see that the linewidth of the injection-locked laser (red curve) is significantly compressed compared to that of unlocked (black curve). The unlocked curve gives a 3-dB linewidth about 5.20 MHz by Gaussian fitting, as shown in Fig. 2(b), while the locked curve only gives a 3-dB linewidth of 1.25 kHz, as shown in Fig. 2(c). In order to give the single-laser-linewidth, the beat signal between the two NPRO lasers is also characterized, which gives a beat signal linewidth of 1.24 kHz (not shown here). Since the two NPRO lasers are identically designed and built, their single-laser-linewidth is thus believed to be identical, which is determined to be 876 Hz. Then, the single-laser-linewidth of the DFB-LD is determined to be 891 Hz after injection locking. In the experiment, the output power of the DFB-LD was set to 100 mW, while the injected power from NPRO1 was set to 10 mW.

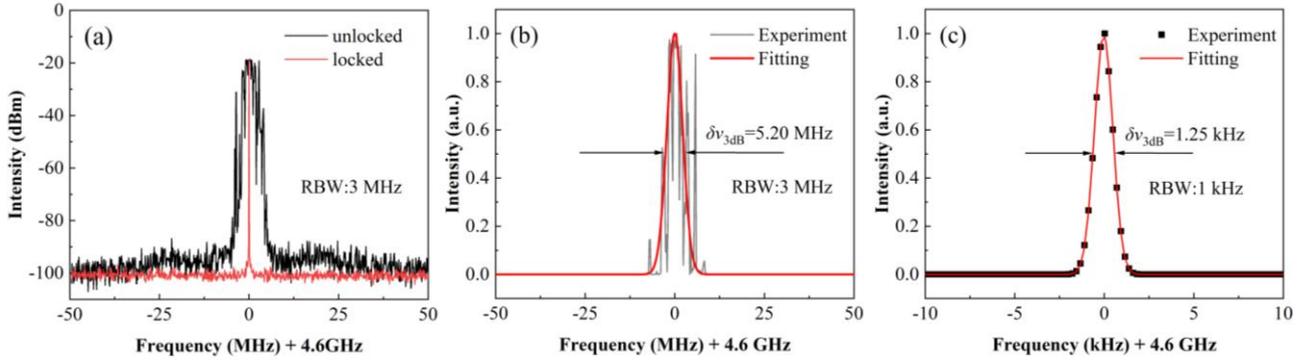

**Fig. 2 (a) Beat signals generated by NPRO2 and the continuously operated DFB-LD under injection-locked and unlocked conditions, respectively. (b) Gaussian fitting of the beat signal under unlocked condition. (c) Gaussian fitting of the beat signal under injection-locked condition. RBW, resolution bandwidth.**

Fig. 3 shows the measurement results of the frequency noise of the beat signals. Here, the black curve represents the frequency noise of the beat signal between NPRO1 and NPRO2, while the blue curve represents the frequency noise of the beat signal between the injection-locked DFB-LD and NPRO2. One can see that these two curves overlap with each other in most frequency range, which means the injection-locked DFB-LD has followed the frequency noise of the NPRO laser. A slight difference occurs around the relaxation oscillation frequency (about 60 kHz) of NPRO laser, where the frequency noise of the DFB-LD is a little bit higher than that of the NPRO laser. We attribute this to the amplifying effect during the injection locking. Note that due to the mode-instability of the DFB-LD under free-running condition, the phase noise analyzer cannot capture the frequency noise of its beat signal with NPRO2. The peak of frequency noise observed at 50 Hz is attributed to the mains frequency. We integrated the frequency noise beyond the Beta separation line (red solid line) [36], commencing at 10 Hz, and determined the beat signal linewidths to be about 1.31 kHz (black curve) and about 1.32 kHz (blue curve), respectively. This finding aligns with the conclusions illustrated in Fig.2, and further proves that the single-laser-linewidth is sub-kHz for both of our NPRO laser and the injection-locked DFB-LD.

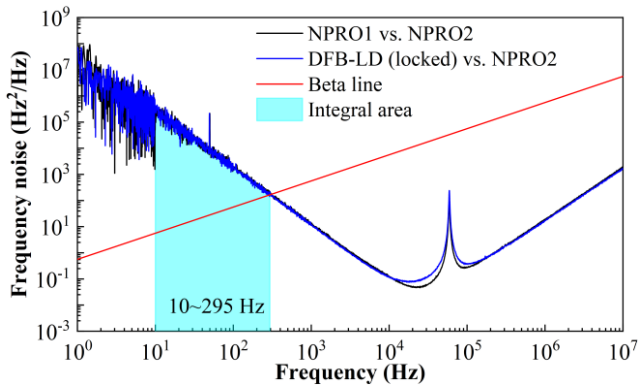

**Fig. 3 Frequency noise curves of the beat signals between NPRO1 and NPRO2 (black curve), and between the injection-locked DFB-LD and NPRO2 (blue curve), respectively. The Beta line (red line) and the linewidth integral area (cyan zone) are also given.**

The output power stability of the DFB-LD laser under injection-locking was tested using an optical power meter (Thorlabs, PM400/SC130C). As shown in Fig. 4, during a 14-hour test, the average output power of the laser is 100.73 mW and the relative power fluctuation is approximately 0.3% root mean square (RMS).

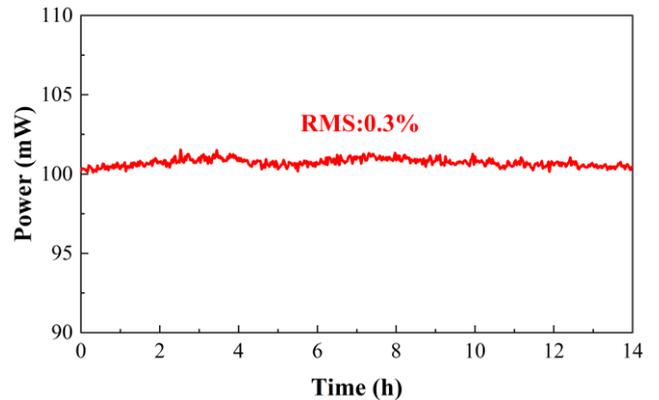

**Fig. 4 Stability test of the output power of the injection-locked DFB-LD in the CW mode.**

Next, we injected the seed laser into the pulse-operated DFB-LD to assess the impact of this technique on pulsed laser performance. The DFB-LD can be easily set up for pulsed operation by modulating its driving current using a function generator. In our measurements, we used square waves to modulate the currents. Fig. 5 shows typical temporal measurement results, where the modulation pulse widths are set at 18 ns (minimum value from the function generator used in the experiment). Four pulse traces with repetition rates of 250 kHz, 500 kHz, 750 kHz, and 1000 kHz are shown in Fig. 5(a), whose peak-to-peak amplitude fluctuations are below 3% regardless of the repetition rates. Fig. 5(b) provides a detailed view of the pulse profile with the minimum pulse duration at different repetition rates, revealing that there is no significant variation in the pulse shape across the different modulation frequencies. It is worth noting that the pulse duration of the laser is approximately 120 ns, which is longer than the duration of the modulation signal. The cause of this phenomenon is that the current driving circuit we are using has longer rise and fall times, thereby limiting the minimum width of the laser pump current pulse. Moreover, the asymmetry of pulses is also attributed to the different rise and fall times under such short pulse condition. Fig. 5(c) demonstrates the peak power at different repetition rates under the condition of minimum pulse duration. At repetition rates ranging from 100 kHz to 700 kHz, constrained by the maximum operating current of the DFB-LD, the peak power output is approximately 160 mW. However, beyond 800 kHz, a declining trend in peak power is observed. This means the modulation bandwidth of the current driving circuit is nearly reached.

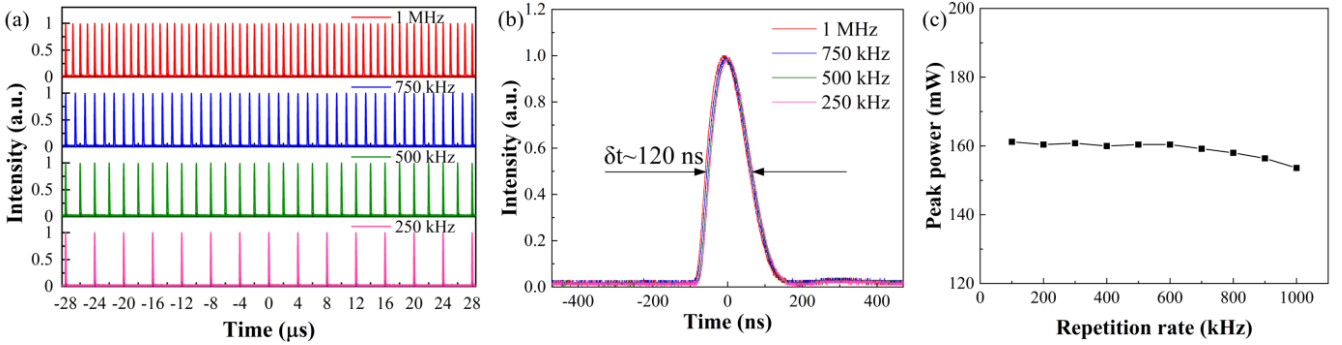

**Fig. 5 (a) Pulse trains under different repetition rates. (b) Pulse shapes under different repetition rates. (c) Peak power as a function of repetition rate.**

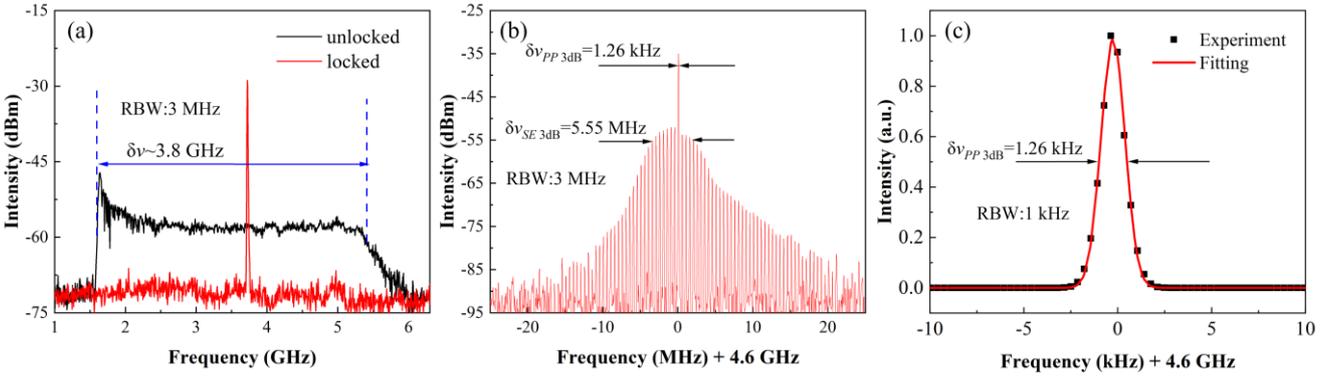

**Fig. 6 (a) Beat signals generated by NPRO2 and the pulse-operated DFB-LD under injection-locked and unlocked conditions, respectively. (b) Detailed lineshape of the beat signal under injection-locked condition. (c) Gaussian fitting of the principal peak in (b). PP, principal peak; SE, sidebands envelope.**

Then, we assessed the laser linewidth of the pulsed laser using the same heterodyne beat frequency method, as shown in Fig. 6. The peak power, pulse duration, and pulse repetition frequency of the DFB-LD were set to 100 mW, 120 ns, and 500 kHz, respectively. Fig. 6(a) shows the beat signal of the pulse-operated DFB-LD and NPRO2. Without injection locking, pulse modulation resulted in an increased linewidth of the DFB-LD, which is about 3.8 GHz as depicted by the black curve in Fig. 6(a). In contrast, following the injection locking to NPRO1, a notable compression of the beat signal spectrum is observed, which is presented by the red curve in Fig. 6(a). Fig. 6(b) shows the details of the beat signal spectrum under injection locking. One can see that a principal peak is accompanied by a series of sidebands that are spaced 500 kHz apart, corresponding to the modulation frequency resulting from intensity modulation. The principal peak has a 3-dB linewidth of 1.26 kHz from Gaussian fitting, as shown in Fig. 6(c). This indicates a single-laser-linewidth of 905 Hz of the pulsed DFB-LD, which is consistent with the linewidth observed in the CW mode (891 Hz). Furthermore, the 3-dB linewidth of the envelope formed by the modulated sidebands is about 5.55 MHz. Notably, both the principal peak and the envelope of the modulated sidebands exhibit significant linewidth compression when compared to the unlocked state.

The frequency noise characteristics of the injection-locked, pulse-operated DFB-LD were also evaluated, and the results are presented in Fig. 7. One can see that the frequency noise curve of the pulsed operation follows closely that of the continuous operation except the designated modulation peaks ($f_N = N \times 500$ kHz, N = 1, 2, 3...). The linewidths of the beat signals are approximately 1.34 kHz and 4.97 MHz, depending on whether the integration of exceeded modulation peaks is included, as determined by the Beta-separation line method. These results are consistent to the 3-dB linewidth of 1.26 kHz and 5.55 MHz in Fig. 6(b). Thus, it further proves the validity of linewidth measurements.

During the frequency noise measurements, we have found that when the repetition frequency is below 50 kHz, the frequency noise will increase significantly relative to that of continuous operation, and the pulsed DFB-LD may lose injection locking. The underlying reasons of this phenomenon require further investigation in the future. Finally, the single-frequency character of the pulsed DFB-LD is verified using a scanning Fabry-Pérot interferometer and an optical spectrum analyzer, as shown in Fig. 8. From Fig. 8(a), one can see two longitudinal modes in the scanning Fabry-Pérot interferometer, which come from two adjacent longitudinal modes with a frequency separation of one free-spectra-range (10 GHz) of the Fabry-Pérot cavity. It thus proves the injection-locked, pulse-operated DFB-LD is working under single frequency condition. Fig. 8(b) further shows its wavelength is 1064.45 nm with a signal-to-noise ratio of 65 dB.

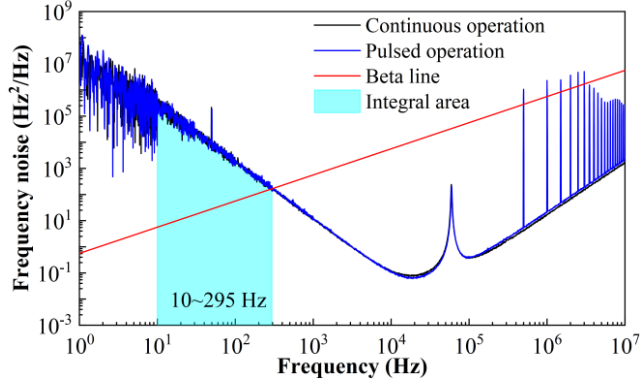

**Fig. 7 Frequency noise curves of the beat signals between injection-locked DFB-LD and NPRO2 under continuous and pulsed operational conditions, respectively. The Beta line (red line) and the linewidth integral area (cyan zone) are also given.**

Table 1 summarizes typical results of recent advancements in SFPLs with different methods and lasing wavelengths. One can see that only this work has demonstrated a SFPL with sub-kHz laser linewidth to our best knowledge. Additionally, our homemade NPRO lasers have two frequency-tuning channels, with tuning coefficient of -3.0 GHz/°C and tuning range of 20 GHz by temperature, and with tuning coefficient of 2.5 MHz/V and tuning bandwidth of 100 kHz by piezo [35,37]. These capabilities will help the single-frequency pulsed DFB-LD has more functions in applications such as frequency-sweep LIDARs [38]. Moreover, due to the low optical power consumption (10 mW) of injection locking, the NPRO laser power (> 100 mW) can be divided into multi-branches to inject multiple DFB-LDs. These multiple, highly coherent, single-frequency pulsed DFB-LDs with same or different repetition frequencies may be used as a distributed measuring system for multidimensional, multiparameter detection [39]. Their optical power can be enhanced by MOPA configurations [37] to extend the detection range.

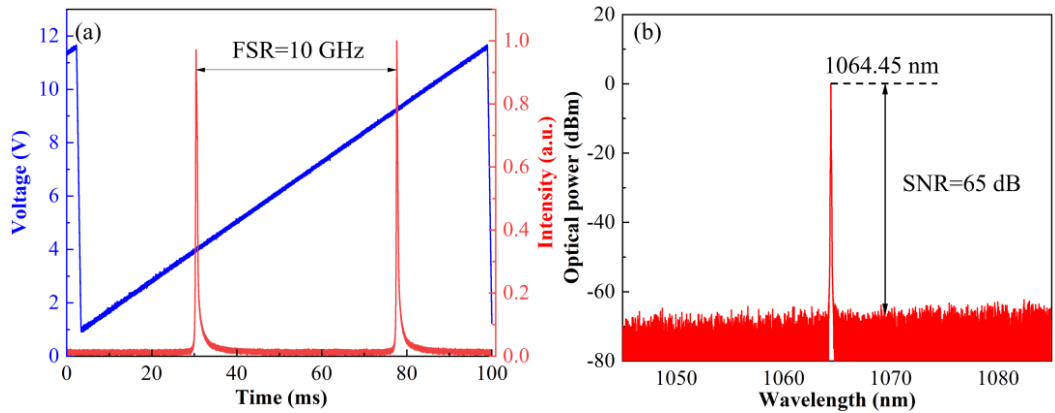

**Fig. 8 (a) Single-frequency performance of the pulse-operated DFB-LD measured by SFPI (FSR 10 GHz, resolution 67 MHz). (b) Optical spectrum of the pulse-operated DFB-LD. The resolution bandwidth is 0.01 nm.**

Table 1. Performance comparisons of SFPLs with different methods and lasing wavelengths.

| Method | Wavelength | Linewidth | Repetition rate | Pulse duration | Ref. |
|---|---|---|---|---|---|
| Direct modulation | 1064 nm | 63 MHz | 20 kHz | 8 ns | [11] |
| | 1064 nm | 283.8 MHz | 80 kHz | 3.8 ns | [40] |
| Q-switch | 1550 nm | 212 kHz | 63.2 - 68.9 kHz | 1.49 - 2.54 µs | [20] |
| | 1645 nm | 1.77 MHz | 100 - 500 Hz | 342.5 - 972.7 ns | [30] |
| Gain-switch | 1063 nm | 14 MHz | 10 - 400 kHz | 150 ns | [22] |
| | 1950 nm | 16.7 MHz | 1 - 500 kHz | 19 - 106 ns | [24] |
| **This work** | **1064 nm** | **905 Hz** | **50 kHz - 1 MHz** | **120 ns - quasi-continuous** | / |

## 4. Conclusion

In summary, a sub-kHz linewidth SFPL has been experimentally demonstrated by directly modulating the drive current of a DFB-LD that is injection locked to an NPRO seed. In the CW mode, the DFB-LD has achieved an extremely narrow linewidth of 873 Hz and optical power of 100 mW at 1064 nm after injection locking is built. In the pulsed mode, the DFB-LD has pulse duration ranging from 120 ns to quasi-continuous state with repetition rates varying between 50 kHz and 1 MHz. The linewidth, maximum peak power and optical spectrum SNR of the SFPL have reached 905 Hz, 160 mW and 65 dB, respectively. At present, the minimum pulse duration and maximum repetition rates are limited by the current driving circuit of the DFB-LD. The maximum peak power is limited by the DFB-LD itself which can be boosted with the help of YDFA in the future. Our results may enhance the methodologies for generating single-frequency narrow-linewidth pulsed seed lasers and are expected to have applications in coherent LIDAR, nonlinear frequency conversion, coherent beam combining, and other related domains.

## Funding

This work was supported by National Natural Science Foundation of China (12404489), Fundamental Research Funds for the Central Universities, Sun Yat-sen University (24QNPY162), National Key Research and Development Program of China (2023YFC2205504, 2020YFC2200200), and Major Projects of Basic and Applied Basic Research in Guangdong Province (2019B030302001).

## CRediT authorship contribution statement

## Declaration of competing interest

The authors declare that they have no known competing financial interests or personal relationships that could have appeared to influence the work reported in this paper.

## Data availability

Data will be made available on request.